# Stress engineering with silicon nitride stressors for Ge-on-Si lasers

**Jiaxin Ke[1], Lukas Chrostowski [2], Guangrui (Maggie) Xia[1]**

[1]Department of Materials Engineering, the University of British Columbia, Vancouver, Canada
[2]Department of Electrical and Computer Engineering, the University of British Columbia, Vancouver, Canada

*gxia@mail.ubc.ca*

**Abstract:** Side and top silicon nitride stressors were proposed and shown to be effective in reducing the threshold current $I_{th}$ and in improving the wall-plug efficiency $\eta_{wp}$ of Ge-on-Si lasers. Side stressors only turned out to be a more efficient way to increase $\eta_{wp}$ than using the top and side stressors together. With the side stressors and geometry optimizations, a $\eta_{wp}$ of 30.5% and an $I_{th}$ of 50 mA ($J_{th}$ of 37 kA/cm$^2$) can be achieved with a defect limited carrier lifetime ($\tau_{p,n}$) of 1 nsec. With $\tau_{p,n} = 10\ ns$, an $I_{th}$ of 7.8 mA ($J_{th}$ of 5.8 kA/cm$^2$) and a $\eta_{wp}$ of 38.7% can be achieved. These are tremendous improvements from the case without any stressors. These results give strong support to the Ge-on-Si laser technology and provide an effective way to improve the Ge laser performance.

## 1. Background and Introduction

Optical interconnects are highly desired for on-chip and short-reach data communications to reduce the resistance-capacitor (RC) delay time and the power consumption. For this purpose, on-chip silicon (Si)-compatible light sources have long been pursued as electrical to optical signal converters, which are important and indispensable components of Si photonics. III-V semiconductor-based lasers integrated on Si via wafer bonding have provided the best performance so far, but they have disadvantages of high cost, low yield, and low integration density, which are not suitable for mass production. The direct hetero-epitaxial growth of III-V materials on Si such as the InAs/GaAs quantum dots (QD) lasers demonstrated in [1] are more promising for low cost, high yield fabrication in the future. However, due to contamination issues, it will take a long time for III-V semiconductors to enter the mainstream Si fabrication facilities (fabs). Ge-on-Si laser is another competitive solution for the large-scale monolithic integration because it is fully compatible with the complementary metal-oixde-semiconductor field effect transistor (CMOS) technology, which may greatly reduce the process complexity, cost and time to enter the fabs [2]. Light emission from Ge by band engineering with tensile strains and high doping levels was theoretically predicted in 2007 [3]. The first optical pumped [4] Ge laser was first realized in 2010 and electrically pumped Ge lasers were demonstrated in 2012 [5] and 2015 [6] accordingly. Other types of Ge lasers like GeSn lasers [7], Ge QD lasers [8] have been demonstrated recently, which show the potential of Ge as a lasing material on Si. Unfortunately, the demonstrated Ge lasers suffer from high threshold current and low efficiencies. The electrically pumped laser in [5] has a threshold current density ($J_{th}$) of 280 kA/cm$^2$, which is too high for any useful applications. Optimization of Ge lasers is in great need to lower the threshold current ($I_{th}$) and increase the efficiency. Bandgap engineering by stress is a very promising way to increase the gain [9] compared with higher doping because high doping introduces high optical loss. Introducing tensile strain to Ge can transfer Ge from an indirect bandgap material into a direct bandgap material and thus increase the gain [10,11]. Both biaxial and uniaxial tensile strain can make this transition. Many efforts have been invested to increase the tensile strain in Ge. D. S. Sukhdeo et al. used a stress concentration method in Ge-on-insulator (GOI) substrates, and obtained 5.7% uniaxial tensile stress in Ge bridges [12]. G. Capellini et al. used silicon nitride layer to stress Ge up to about 1.5% uniaxial tensile strain, and the fabrication process was CMOS-compatible [13]. Simulation results showed that silicon nitride (SiN) top stressor could reduce $I_{th}$ by 2 to 3 times [14]. A highly stressed Ge photodetector has been achieved to reach a detection range up to 1.8 µm [15].

Our previous simulation work showed that by adjusting the geometry of Ge cavity and increasing the cladding thickness, Ge laser's performance could be significantly improved [16]. To further enhance Ge laser performance, we used SiN stressors to introduce tensile strain in this work. Silicon nitride has been widely used in CMOS industry to introduce both tensile and compressive stress. The stress levels of SiN can be easily tuned by changing the deposition recipe. Intrinsic stress values of ±2Gpa were used in our simulations, which were achievable in CMOS technology [17]. SiN has a refractive index about 1.6 to 2. Therefore, it is suitable for the optical confinement too. Based on the MIT's experimental laser structure in [5], we proposed side and top nitride stressors to introduce stress in the Ge cavity. Three double-heterojunction Fabry-Perot laser structures were simulated to study the stressors' impact on the device performance (Fig. 1). Structure 1 is the simplified version of the experimental laser structure in [5].

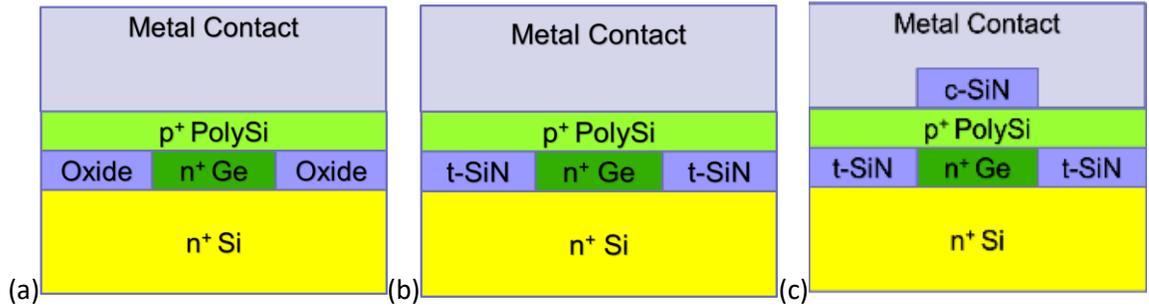

Fig. 1. Laser structure simulated (cavity width = 1 µm, thickness = 0.2 µm, length = 270 µm, cladding thickness = 0.18 µm). (a) Structure 1: without stressors (b) Structure 2: with side nitride stressors (c) Structure 3: with top & side nitride stressors. The width of the top nitride stressor is the same as that of the Ge cavity. The metal contacts are composed of Ti and Al same as those in Ref. [5], shown in Fig. 2 (a).

## 2. Laser structures, parameters used and calibration

Laser structures simulated in this work use MIT's experimental structure [5] as a start point. The cross section is illustrated in Fig. 2. The doping and the strain are the same as the experiments reported: Si substrate is $5\times10^{19}$ cm$^{-3}$ n-type doped; Ge is $4\times10^{19}$ cm$^{-3}$ n-type doped with 0.25% biaxial tensile strain; poly-Si is $5\times10^{20}$ cm$^{-3}$ p-type doped. 2 µm Si substrate was used in the simulations. A virtual contact was defined underneath the bottom of the Si substrate and the top of metal layers for the biasing purpose. The structure was 1µm wide and 270 µm long with 180 nm thick poly-Si cladding layer. The thickness of Ge active layer was set to be 200 nm, which was the average value of the 100~300 nm thickness in the experiments due to the process non-uniformity [5,18].

The strain-dependent Ge energy bandgap model in Refs. [19, 20] and the doping induced bandgap narrowing effect [21] were implemented in a commercial 2D laser simulation tool LASTIP$^{TM}$. The metal-semiconductor heterojunctions were aligned by electron affinity as described in [27]. The reflectivity values of two facet are $R_1$ = 23% and $R_2$ = 38%, which correspond to a mirror loss $\alpha_m$ of 45 cm$^{-1}$ [22]. Auger coefficients used were $C_{nnp}$= 3.0× 10$^{-32}$cm$^6$/s and $C_{ppn}$= 7.0× 10$^{-32}$cm$^6$/s [23]. The index of refraction values of all materials were wavelength dependent. The material parameters used mainly come from Ref [5,16]. 1 ns of defect limited carrier lifetime ($\tau_{p,n}$) was used as a conservative estimation [24].

For the optical loss, we assumed that the internal loss and mirror loss is the primary sources of the loss and internal loss is dominated by the free carrier absorption [25]. In LASTIP$^{TM}$, for a narrow wavelength range, the free carrier absorption is described by $\alpha_i = AN + BP$, where A, B are constants and N, P are the electron and hole density in the unit of cm$^{-3}$. We used the first principle calculations results of free carrier absorption in n-type doped Ge for n-loss coefficient $A = 5.0 \times 10^{-19}$ [18] and the experimental measurement results in p-type doped Ge [26] as a starting point to obtain the best fitting to the L-I curve in [5]. The effective mass of gamma conduction band ($m_{e\Gamma}$*)was used as the first fitting parameter of L-I curve since it usually deviates from the typical measurement result $m_{e\Gamma}$* =0.038$m_e$. This deviation may come from the temperature difference between the measuring temperature and the operating temperature as well as the interdiffusion between Si and Ge that changes some portion of Ge into a SiGe alloy. The p-loss coefficient B was used as the second fitting parameter. The best fitting was obtain when $m_{e\Gamma}$* = 0.045735$m_e$ and the best fitting free carrier loss relation was $\alpha_i = 5.0 \times 10^{-19}N + 1.023 \times 10^{-17}P$.

Using these parameters, our model produced $J_{th}$ of 300 kA/cm$^2$ or $I_{th}$ of 800 mA at 15°C with the transverse electric (TE) mode lasing at λ= 1676 nm, which were very close to the experimental values of $J_{th}$ = 280 kA/cm$^2$ and lasing wavelength range of 1650 nm [5]. As seen in Fig. 3, the model could match the experimental L-I curve quite well. Sensitivity test results are shown in Fig. 3, which shows how a smaller FCA parameter or a smaller $m_{e\Gamma}$* are not fitting the experimental data. After the calibration of our model, we started optimizing the laser structure.

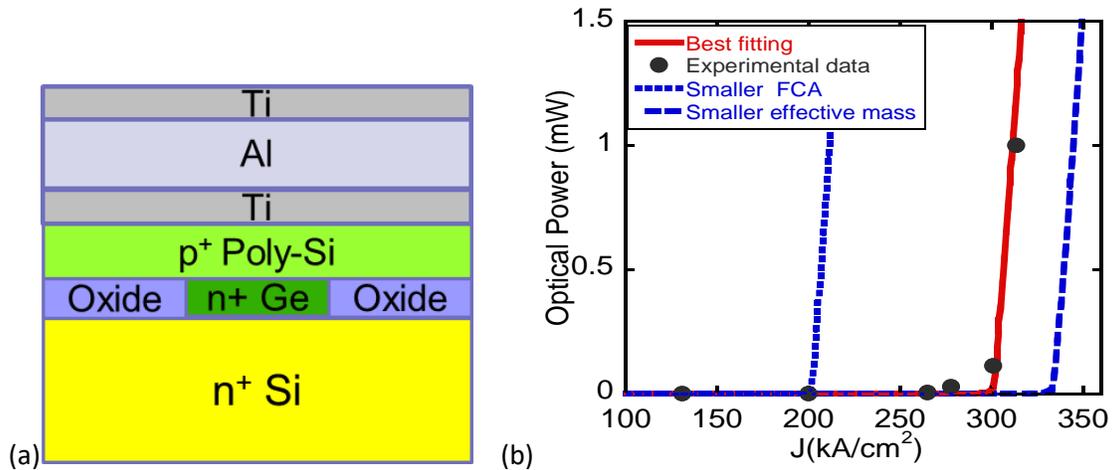

Fig. 2. (a) Cross-section of the Ge-on-Si heterojunction laser structure simulated. (b) L-I curves for experimental result, calibration result, sensitivity tests with a smaller FCA coefficient for holes: $\alpha_i = 5.0 \times 10^{-19}N + 0.923 \times 10^{-17}P$, and a smaller effective mass $m_{e\Gamma}$* = 0.045335 $m_e$.

## 3. Stress modeling and impact on laser performance

The stress simulations were performed using a standard two-dimensional (2D) process simulation tool TSUPREM-4$^{TM}$. 2D stress/strain simulations are suitable for Ge-on-Si lasers as the length dimension is much longer than the width and thickness dimensions. The intrinsic stress values used for tensile silicon nitride (t-SiN) and compressive silicon nitride (c-SiN) were +2 and -2 GPa respectively. The strain was calculated from the model in [27]. For each geometry, the stress was calculated by TSUPREM-4 including a biaxial thermal mismatch strain of 0.25% due to the thermal expansion mismatch between Ge and Si. As LASTIP$^{TM}$ was only able to model the energy band structure under biaxial strain conditions, some approximations were needed. As the lateral (σ$_{xx}$) and the longitudinal (σ$_{yy}$) stresses are close (Fig. 3(a)) and the vertical stress is much smaller, we could approximate the stress status with an effective biaxial stress defined as $\sigma_{eb} = \frac{(\sigma_{xx}+\sigma_{yy})}{2}$. The effective biaxial strain $\varepsilon_{eb} = \frac{(\varepsilon_{xx}+\varepsilon_{yy})}{2}$ and the vertical strain $\varepsilon_{zz}$ were then calculated from Hook's law using the compliance matrix of Ge and were then loaded into LASTIP$^{TM}$ for device simulations.

Structure 2 and 3 have the same cavity sizes as those in MIT's experiments but with the SiN stressors as illustrated in Fig. 1(b) and (c). Significant lateral and longitudinal stresses were introduced by the stressors (Fig. 3(b)). With only side stressors, 0.4% $\varepsilon_{eb}$ was introduced including the 0.25% strain caused by the thermal expansion mismatch. The value was increased to 0.53% by adding a top stressor. This strain enhancement is not optimized due to the non-optimized Ge width and thickness as the stress introduction strongly depends on the stressor and the cavity's sizes and relative positions. Compared to Structure 1, by adding the side stressors, about 590 mA reduction in $I_{th}$ and 1.2% increase in $\eta_{wp}$ were obtained. By adding the top and side stressors, about 760 mA reduction in $I_{th}$ and 13.6% growth in $\eta_{wp}$ were obtained (Fig. 4 and Table.1). The significant performance improvement introduced by the top stressors is because that top stressor not only introduces higher stress, but also decreases the optical loss caused by the metal contact and provides optical confinement in the vertical direction.

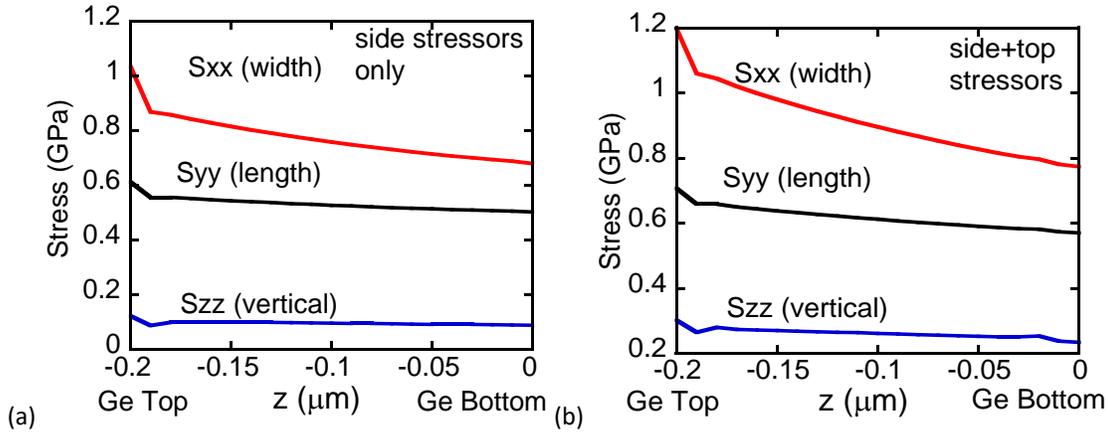

Fig. 3. 1D vertical cut of the stress at the center line of the Ge cavity: (a) with side stressors only and (b) with side and top stressors

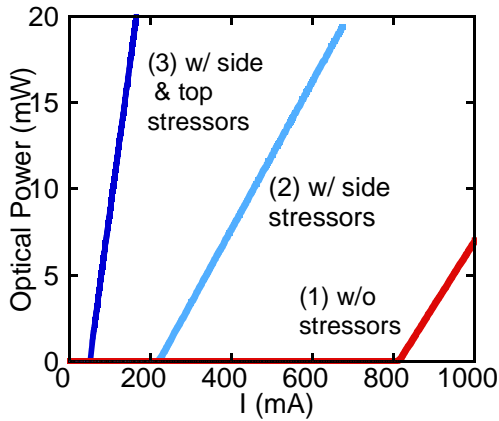

Fig. 4. L-I curve comparison for the three structures in Fig. 1 before the structure optimizations.

| Structure | 1 | 2 | 3 |
|---|---|---|---|
| $\varepsilon_{eb}$ | 0.25% | 0.4% | 0.53% |
| $I_{th}$ (mA) | 810 | 220 | 50 |
| $\eta_d$ | 6.33% | 7.50% | 31.34% |
| highest achievable $\eta_{wp}$ | 2.07% | 3.53% | 15.03% |

Table.1 Laser performance of the three structures in Fig. 4

### 4. Laser structure optimization methodology and Structure 2 optimizations

In order to take the full advantage from the stressors and further improve the device performance, we optimized the Ge cavity and the cladding geometry. We chose W, $d_{Ge}$, and $d_{poly}$ as the parameters to be optimized, which stand for the Ge cavity width, thickness, and the poly-Si cladding layer thickness respectively. The Ge cavity length was set to be unchanged at 270 μm. In our optimization process, the goal is not to find the "true" optimal point, but rather to show that Ge lasers can be improved significantly. The reasons for that are two-fold. 1) Ge is not a well-studied optical material, and many model parameters do not have widely agreed values or even ranges. Therefore, it is still too early to find the "true" optimal at this point. 2) Optimizing one variable at a time is more doable, as the rate equations are well established, and one can check the correctness of the results conveniently.

Laser rate equations below were used to analyze the results. Slope efficiency η$_d$ is calculated from Eq. (1), where ΔP/ΔI is the slope of L-I curve, c is the speed of light and h is the Planck's constant. η$_d$ is the product of the internal efficiency η$_i$ and the extraction efficiency η$_{ext}$. <α$_i$> in Eq. (2) is the weighted average of the local loss. $R_{srh}(n_{th}, p_{th})$ and $R_{Aug}(n_{th}, p_{th})$ are the non-radiative recombination coefficients due to traps and Auger process respectively. $R_{rad}(n_{th}, p_{th})$ is the spontaneous recombination rate. η$_{wp}$ is defined here as the max wall-plug efficiency that can be achieved. 2D laser L-I and I-V simulations were performed up to about 10 mW optical output, above which, to save computation time, L-I and I-V curves were extrapolated linearly up to about 200 mW optical output, based on which the maximum η$_{wp}$ was determined.

$$\eta_d = \frac{\Delta P}{\Delta I} / \frac{hc}{q\lambda} = \eta_i \frac{\alpha_m}{<\alpha_i> + \alpha_m} = \eta_i \eta_{ext} \quad (1)$$

$$\eta_{ext} = \frac{\alpha_m}{<\alpha_i> + \alpha_m} \quad (2)$$

$$I_{th} = \frac{qdW}{\eta_i}\left(R_{srh}(n_{th}, p_{th}) + R_{rad}(n_{th}, p_{th}) + R_{Aug}(n_{th}, p_{th})\right) \quad (3)$$

$$\eta_{wp} = \text{Max}\left[\frac{P_{op}}{I \cdot V}\right] \quad (4)$$

In most of our optimizations, we changed one parameter at a time and kept others unchanged except the case of Structure 3 where both top and side stressors were optimized together (details in Section 5.3). Next, we will use Structure 2 as an example to illustrate the optimization process. For lasers, small I$_{th}$ and large η$_{wp}$ are both desired, but they may not be met at the same time. We chose η$_{wp}$ as the most important optimization criteria because it represents the energy efficiency of the device.

### 4.1 Poly-Si thickness d$_{poly}$ optimizations

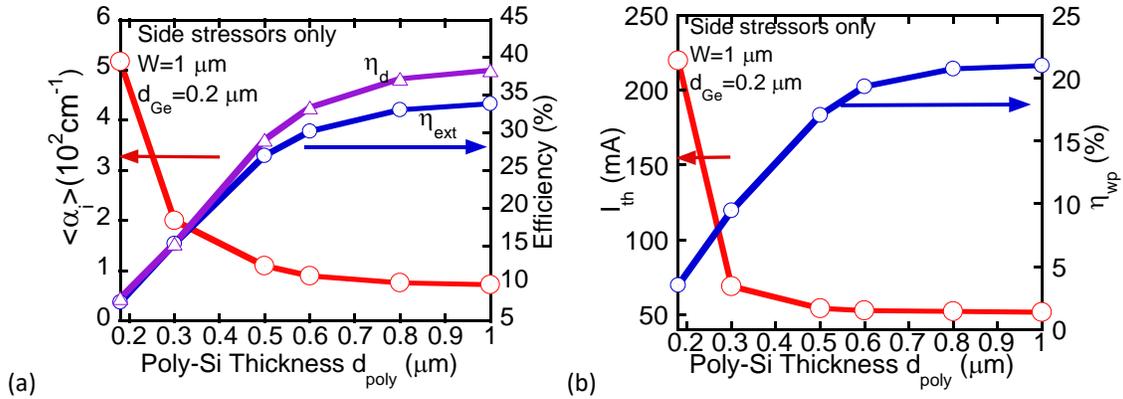

Fig. 5. Poly-Si thickness d$_{poly}$ dependence (W =1 μm, d$_{Ge}$ = 0.2 μm) of (a) <α$_i$> and η$_{ext}$, η$_d$ (b) I$_{th}$ and η$_{wp}$.

The poly-Si thickness d$_{poly}$ has the most dominant effect in the geometry optimization. As d$_{poly}$ increased, we observed a dramatic increase in η$_{wp}$ and a decrease in I$_{th}$. These improvements are because that the metal is very lossy optically. As the top metal contact moved further away from the Ge cavity with the increase in d$_{poly}$, <α$_i$> caused by the metal contact decreases. As a result, η$_{ext}$ and thus η$_d$ increase monotonically and plateau at thick d$_{poly}$ (Fig. 5(a)). I$_{th}$ decreases as d$_{poly}$ increases since less carrier density is needed to compensate for the loss. As a consequence, η$_{wp}$ increases to 20.8% and plateaus after d$_{poly}$ = 0.8 μm and I$_{th}$ decreases to 52 mA (Fig. 5(b)). We chose 0.8 μm as the optimization point since η$_{wp}$ plateaued after that point.

## 4.2 Ge Width W optimizations

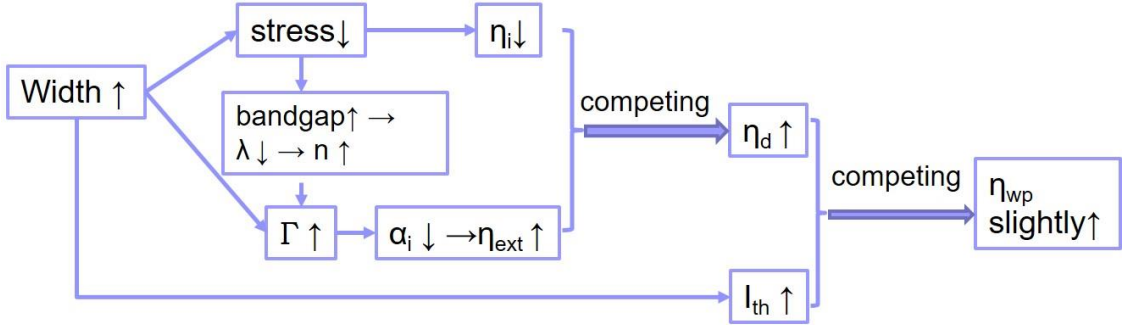

Fig. 6. Impacts of Ge Width (W) to other parameters

The W and d dependence come from three different effects: 1) stress introduction, 2) optical confinement factor Γ and 3) active region volume. The relationship between different parameters is shown in Fig. 6. The tensile stress decreases with the increase of W because side stressors are moved away from the center (Fig. 7(a)). The separation between the direct and indirect band gap increases accordingly, which results in a reduction in $\eta_i$. The decrease in stress raises the bandgap, causing the lasing wavelength to become smaller. The reduced lasing wavelength causes a slight increase in the refractive index and thus increases Γ. As the cavity becomes wider, the lateral confinement becomes better, which also increases Γ (Fig. 7(a)). The FCA loss of poly-Si is bigger than Ge. So a bigger Γ means fewer light travels in the lossy poly-Si region, which results in the decrease of $<\alpha_i>$ and the increase of $\eta_{ext}$ and thus the growth of $\eta_d$ (Fig. 7(b)(c)). $I_{th}$ is a combination effect of $n_{th}$, $\eta_i$ and geometry as indicated in Eq. (3) but mostly dominated by geometry since $I_{th}$ increases almost linearly with W in Fig. 7(d). The bigger the W is, the larger current is needed to compensate the carrier loss resulted mainly from $R_{srh}$ and $R_{Aug}$.

The increase of $\eta_d$ would increase $\eta_{wp}$ whereas increased $I_{th}$ would decrease it. Because of the competing effect, $\eta_{wp}$ only increase slightly with W as shown in Fig. 7(d). Further simulations show that choosing the maximum $\eta_{wp}$ point where W=1μm does not promise better performance in d dependence since a narrower waveguide is desired for side stressors. On the contrary, a wider cavity increases $I_{th}$ greatly. So, we chose W=0.5 μm as the optimization point, where $\eta_{wp}$ =18.21% now but promote better potential for large efficiency.

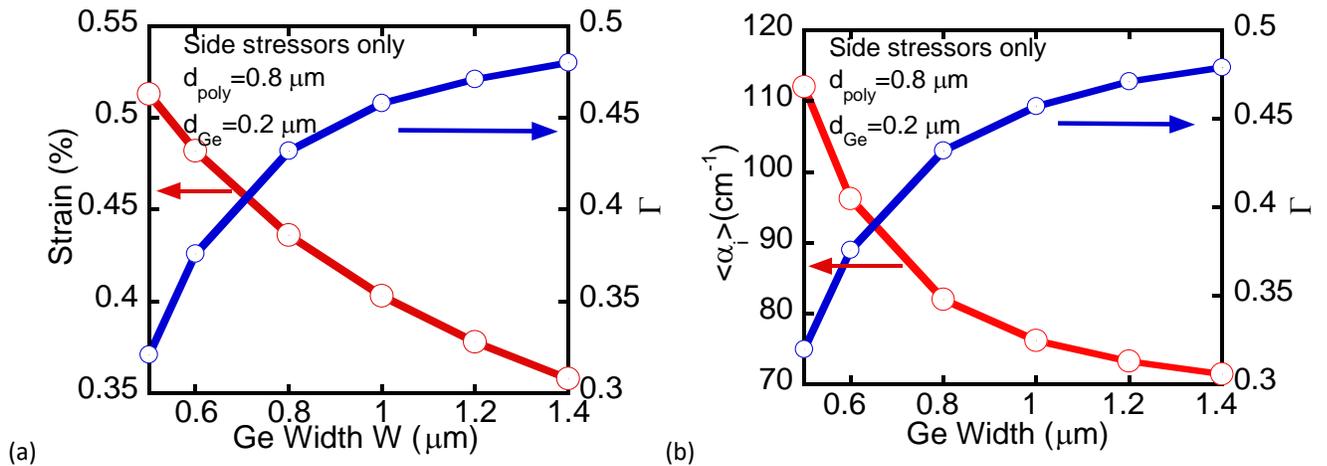

(a) (b)

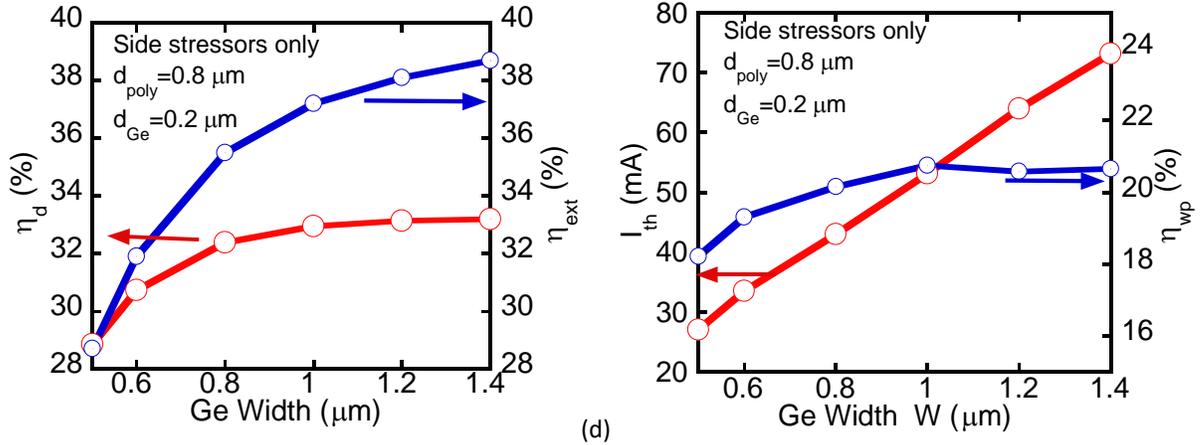

Fig. 7. Ge Width W dependence ($d_{poly}$ =0.8 μm, $d_{Ge}$ = 0.2 μm) of (a) strain and Γ (b) <$α_i$> and Γ (c) $η_d$ and $η_{ext}$ (d) $I_{th}$ and $η_{wp}$.

### 4.3 Ge Thickness $d_{Ge}$ optimizations

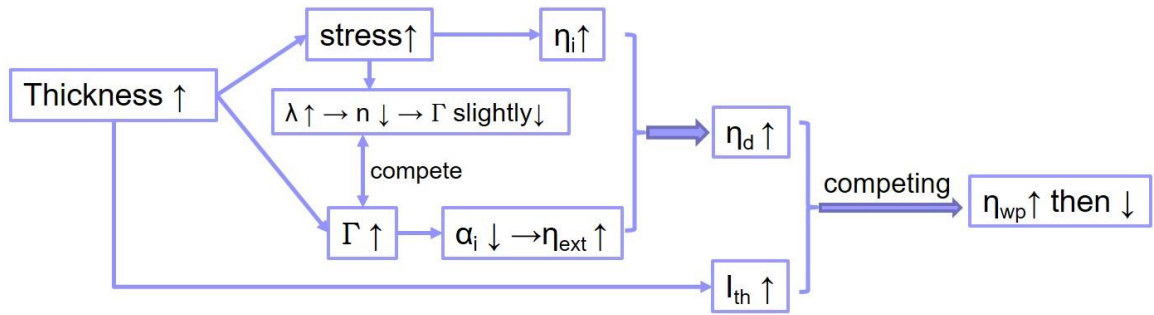

Fig. 8. Impacts of Ge thickness $d_{Ge}$ to other parameters

The dependence of $d_{Ge}$ is similar as W's dependence, which is shown in Fig. 8. Strain and $η_i$ increase with $d_{Ge}$ because more stressors react on the Ge cavity (Fig. 9(a)). Γ increases with $d_{Ge}$ since thicker cavity promote better vertical confinement (Fig. 9(a)). The Γ shrinkage due to the change of lasing wavelength is only a minor effect for Γ. The increase in Γ causing the <$α_i$> to shrink and thus increase the $η_{ext}$(Fig. 9(b)(c)). As a result, $η_d$ increases a lot since $η_i$ and $η_{ext}$ are of same trend (Fig. 9(c)). Same as the W dependence, $I_{th}$ increases almost linearly with $d_{Ge}$ (Fig. 9(d)). As the competing effect of $I_{th}$ and $η_d$, $η_{wp}$ peaks at 30.5% and then decreases (Fig. 9(d)). We chose $d_{Ge}$ = 0.8 μm as the optimization point.

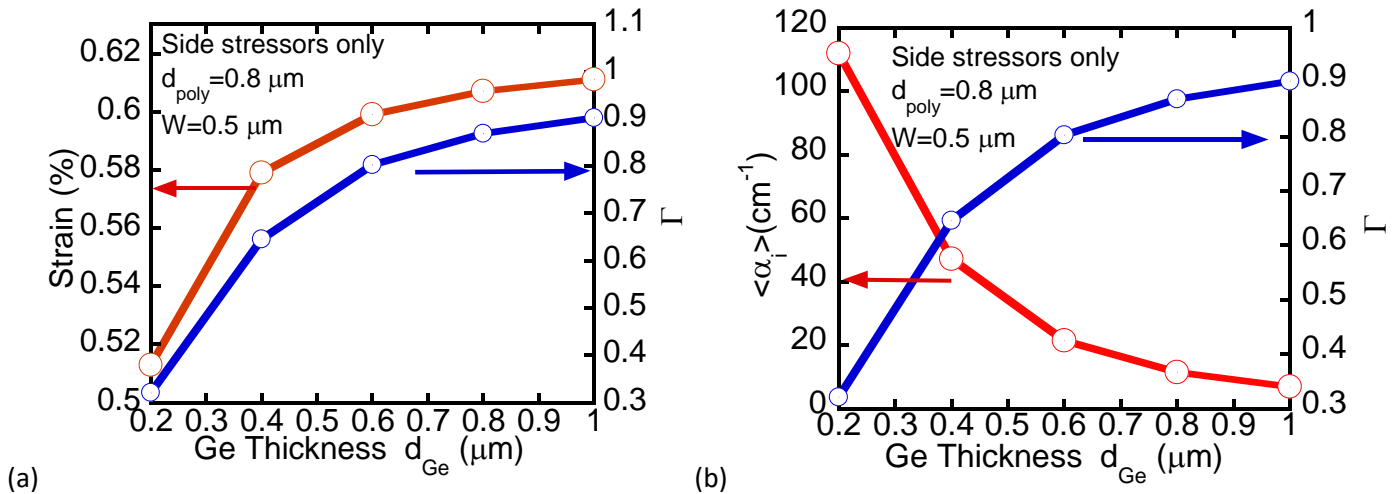

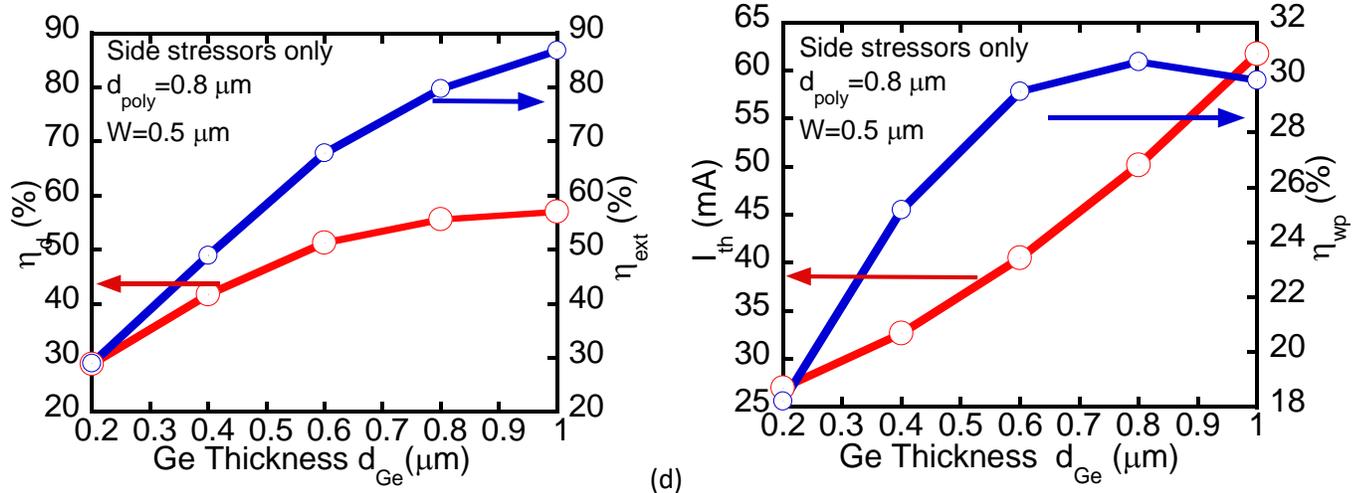

Fig. 9. Ge thickness $d_{Ge}$ dependence (W =0.5 μm, $d_{poly}$ = 0.8 μm) of (a) strain and Γ (b) <$α_i$> and Γ (c) $η_d$ and $η_{ext}$ (d) $I_{th}$ and $η_{wp}$.

By comparing the same structure in $d_{Ge}$ dependence with and without stressors, we can see how stress influences the laser performance. Increased stress decreases the difference between gamma (direct) and L (indirect) conduction band (not shown here), decreases the bandgap and increases the gap between lh and hh band (Fig. 10(a)). These changes in band increase the material gain (Fig. 10(b)), which decreases the carrier density needed for lasing and thus reduce $I_{th}$ (Fig. 10(d)). The increased lasing wavelength (Fig. 10(c)) decreases Γ by the changed real index n and decreases $η_{ext}$ slightly as discussed before (Fig. 10(e)). The $η_d$ increases while $η_{ext}$ decreases by the decreased Γ, which shows that the $η_i$ increases with the stress for the same geometry (Fig. 10(f)).

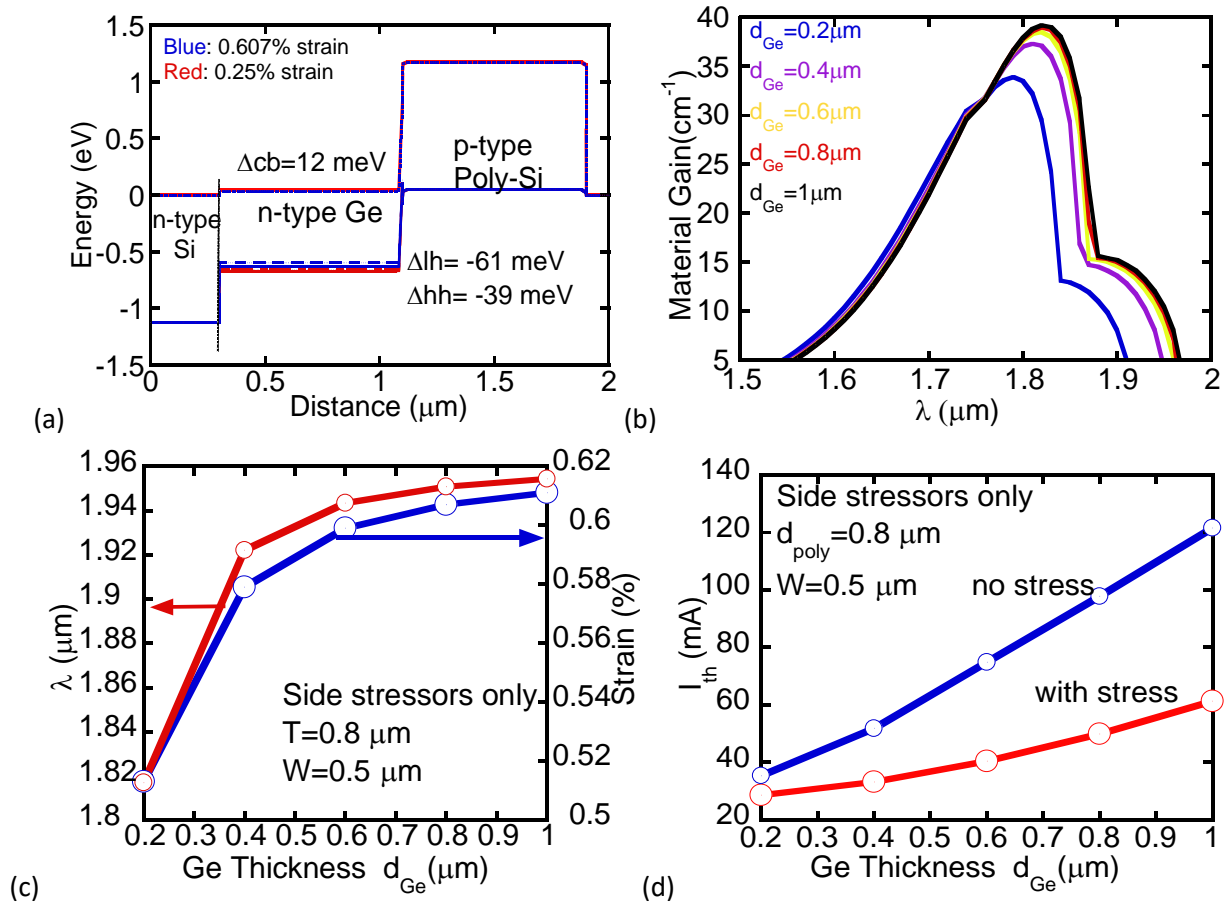

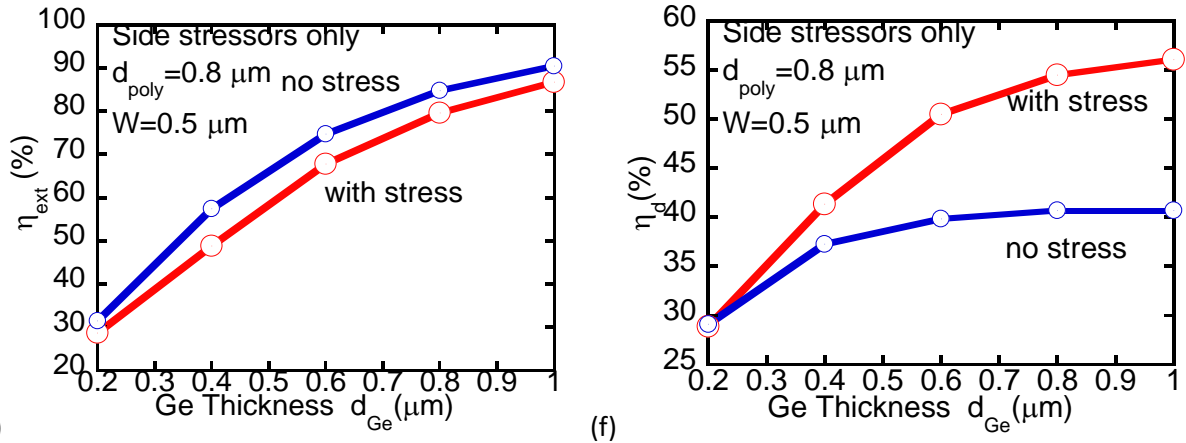

Fig. 10. Stress impact with different $d_{Ge}$ (W=0.5μm, $d_{poly}$ = 0.8μm): (a) direct band alignment under and without strain ($d_{Ge}$=0.8), (b) Material gain at different strain, (c) strain and λ, (d) $\eta_{ext}$, (e) $I_{th}$, and (f) $\eta_d$.

## 5. Structure 1 and 3 optimizations and comparisons between structures

### 5.1 Optimizations of Structure 1 without stressors

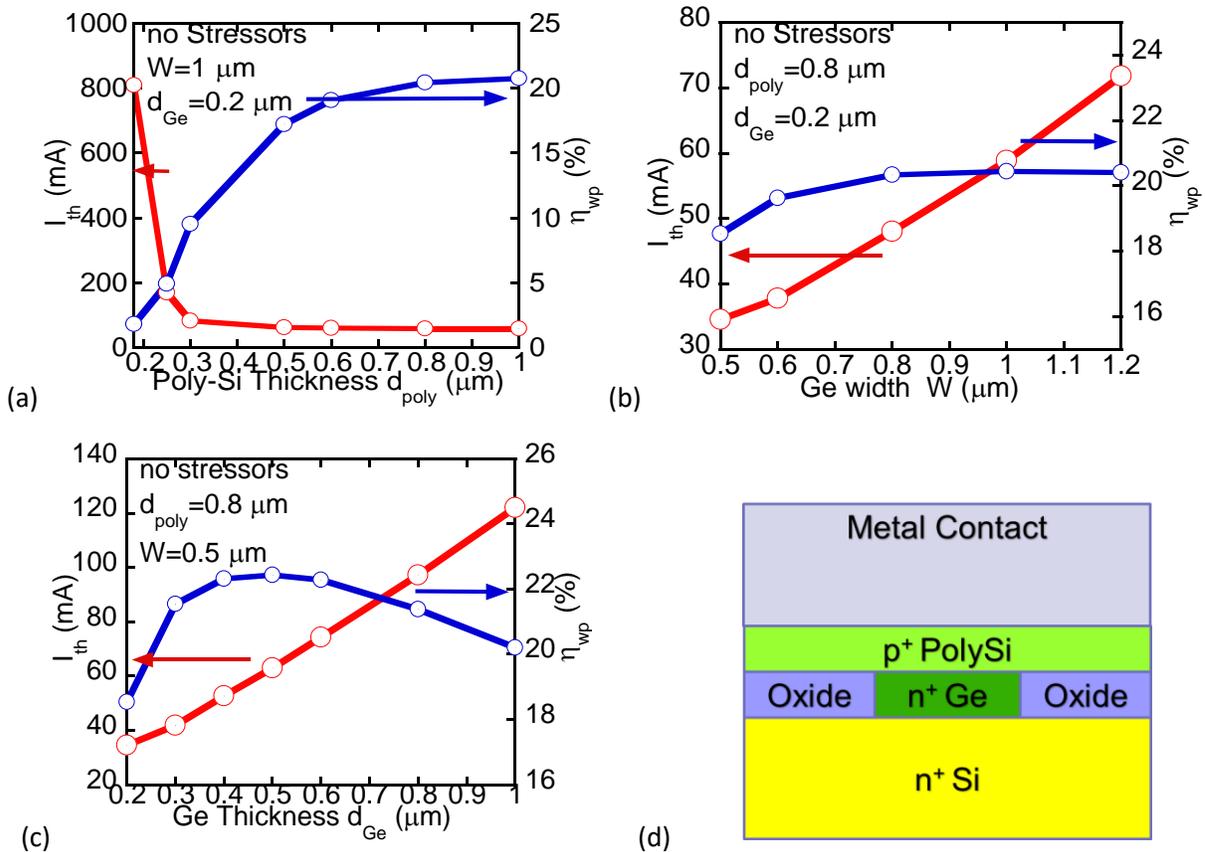

Fig. 11. $I_{th}$ and $\eta_{wp}$ of Structure 2 (a) $d_{poly}$ dependence (b) width dependence (c) $d_{Ge}$ dependence (d) cross section of Structure 1

For Structure 1, without the stressors, the trend is similar to Structure 2 as shown in Fig. 11 but with lower $\eta_{wp}$. $d_{poly}$ had the largest impact and was first optimized as in Fig. 11(a). $I_{th}$ decreases from 810 to 57 mA and $\eta_{wp}$ increases from 2.07% to 20.8% when $d_{poly}$ changes from 0.2 to over 0.8. We chose $d_{poly}$ = 0.8 μm as the optimized $d_{poly}$. For the W dependence, $I_{th}$ increases linearly with W, but $\eta_{wp}$ doesn't change much with W(Fig. 11(b)). So we chose W = 0.5 µm as the optimized W for less $I_{th}$. We chose the peak point $d_{Ge}$ =0.5 µm as the optimization point for $d_{Ge}$ dependence. The highest efficiency reached is 23.5% with $d_{poly}$ = 0.8 µm W = 0.5 µm，$d_{Ge}$ = 0.5 µm $I_{th}$ = 63 mA.

### 5.2 Optimization summary of Structure 2 with side stressors only

Detailed geometry dependence and the optimized process were discussed in 4.1-4.3. After the optimizations, the highest $\eta_{wp}$ achievable is 30.5% with $d_{poly}$ = 0.8 µm W = 0.5 µm, $d_{Ge}$ = 0.8 µm.

### 5.3 Optimizations of Structure 3 with top and side stressors

Due to the presence of the top stressor, the top metal contact loss is greatly reduced for Structure 3 before increasing $d_{poly}$. A large W and small $d_{Ge}$ are desired for the stress introduction from the top stressor, which is undesired for the side stressors. Therefore, W and $d_{Ge}$ are optimized together to obtain a high $\eta_{wp}$. W = 0.5 µm is not the optimized width in Fig. 12 (a), but by comparing a few different W values, W = 0.5 µm has the potential to produce a higher $\eta_{wp}$. $d_{Ge}$ = 0.8 µm is the optimization point of $d_{Ge}$ dependence for W = 0.5 µm. For such structures, the stress introduction from side stressors is more prominent than the top stressor. $d_{poly}$ has similar but weaker impact compared to Structure 2. The increasing $d_{poly}$ would further increase $\eta_{wp}$ and decrease $I_{th}$, which shows that top stressor can only diminish the optical loss caused by metal to a certain extent. Increasing $d_{poly}$ is a more efficient way to reduce the optical loss. The final optimization is: $d_{poly}$ = 0.6 µm, W = 0.5 µm, $d_{Ge}$ = 1 µm, with $\eta_{wp}$ peaks at 24% and an $I_{th}$ of 62 mA.

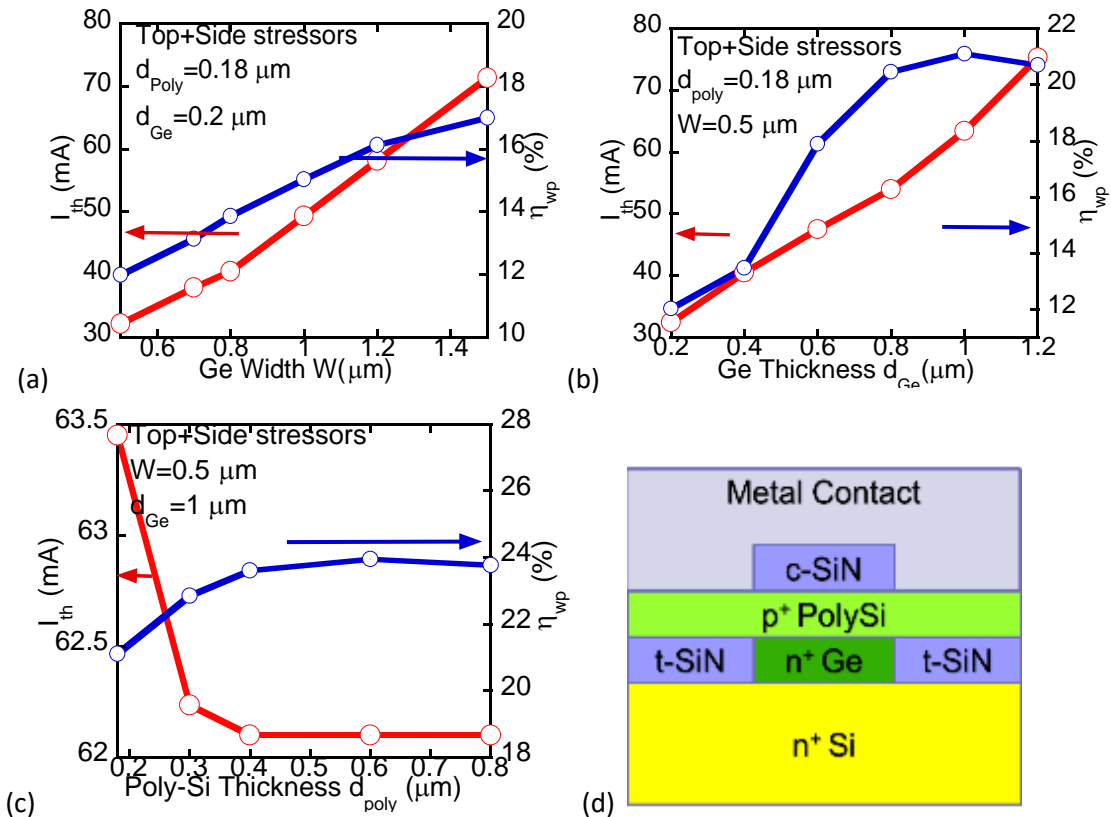

Fig. 12. $I_{th}$ and $\eta_{wp}$ of Structure 3 (a) width dependence (b) $d_{Ge}$ dependence (c) $d_{poly}$ dependence (d) cross section of Structure 3

### 5.4 Comparisons of the structures

The comparisons of the three structures after optimization are shown in Fig. 13 and Table. 2. We can observe that changing geometry could significantly increase η$_{wp}$ and decrease I$_{th}$. Adding stressors can further improve them. By using side stressors only, the highest η$_{wp}$ rose to 30.5%, but adding top stressor does not provide greater η$_{wp}$. This is mainly because that the top stressor increases the series resistance significantly. Structure 1 and 2 have a series resistance around 0.3 Ω, but it is 0.8 Ω for Structure 3, which means Structure 3 requires higher voltage and thus higher electric power. As a result, Structure 3 does not produce a higher η$_{wp}$ than Structure 2. Therefore, considering both η$_{wp}$ and I$_{th}$, Structure 2, with side stressors only, is recommended.

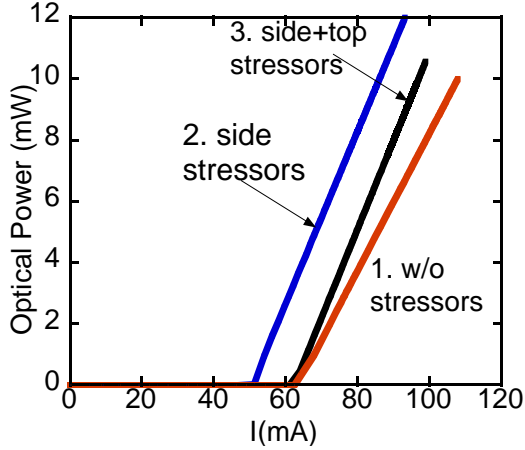

Fig. 13. L-I curve for three structures after optimization

| Structure | 1 | 2 | 3 |
|---|---|---|---|
| ε$_{eb}$ (%) | 0.25 | 0.607 | 0.611 |
| I$_{th}$ (mA) | 63 | 50 | 62 |
| J$_{th}$ (kA/cm$^2$) | 47 | 37 | 46 |
| η$_d$ (%) | 38.6 | 55.5 | 56.0 |
| highest η$_{wp}$ (%) | 23.5 | 30.5 | 24.0 |
| Current required for highest η$_{wp}$ (mA) | 494 | 324 | 258 |
| Output power at highest η$_{wp}$ (mW) | 100 | 78 | 58 |

Table. 2 Laser performance of the 3 structures in Fig. 13 after structure optimizations.

## 6. Defect-limited minority carrier lifetime dependence

For the study above, the defect-limited minority carrier lifetime $\tau_{p,n}$ is set as 1ns for conservative prediction. It is used for Shockley-Read-Hall (SRH) recombination rate calculation, which is defined by: $\tau_{p,n} = \frac{1}{\sigma_{p,n} N_t v_{p,n}}$ in which $v_{p,n}$ is the thermal velocity of hole and electron and $\sigma_{p,n}$ is the effective capture cross-section of the deep defect state traps. For simplicity, a default setting in the software was used: we assumed that $\tau_p$ and $\tau_n$ are the same and a uniform distribution of donor mid-gap traps with a density of 10$^{10}$ m$^{-3}$. $\sigma_{p,n}$ is calculated from the relationship of $\tau_{p,n}$ above and then used in the calculation of SRH recombination rate R$_{SRH}$.

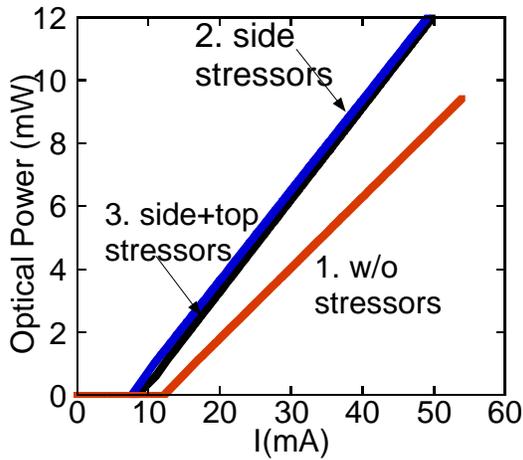

Fig. 14. L-I curve for three structures with $\tau_{p,n} = 10 ns$

| Structure | 1 | 2 | 3 |
|---|---|---|---|
| ε$_{eb}$ (%) | 0.25 | 0.607 | 0.611 |
| I$_{th}$ (mA) | 12 | 7.8 | 8.6 |
| J$_{th}$ (kA/cm$^2$) | 8.8 | 5.8 | 6.3 |
| η$_d$ (%) | 38.7 | 55.8 | 57.0 |
| highest achievable η$_{wp}$ (%) | 27.0 | 38.7 | 35.8 |
| Current required for highest η$_{wp}$ (mA) | 167 | 121 | 73 |
| Output power at highest η$_{wp}$ (mW) | 35 | 33 | 19 |

Table 3 Laser performance of the three structures in Fig. 14

Fig. 14 shows the performance of the 3 structures with $\tau_{p,n} = 10\ ns$. From the results, we can see that by improving the material quality, the performance of laser could improve greatly. $I_{th}$ decreases about 5-8 times when the defect limited carrier lifetime $\tau_{p,n}$ increases from 1ns to 10ns. Longer carrier lifetime means carrier decays slower in the cavity and thus less injection carrier is needed for lasing, which reduces $I_{th}$ accordingly. Although it doesn't change the highest η$_{wp}$ significantly, it decreases the current needed to reach that point significantly. Technically, it is feasible to obtain Ge layers with better quality and longer carrier lifetime by approaches like Ge growth on a GOI substrate [28] or direct wafer bonding and chemical mechanical polishing (CMP) [29]. Defect-limited minority carrier lifetimes of 5.3 and 3.12 ns have been achieved respectively by the above approaches [28, 29]. Therefore, if a better material quality can be achieved, along with the geometry and stress engineering, Ge laser performance will not be too far from III-V lasers.

## 7. Conclusions

We implemented the strain and doping dependent Ge energy bandgap model in LASTIP$^{TM}$ and studied the stress engineering of Ge-on-Si lasers using the silicon nitride stressors. Side and top silicon nitride stressors were proposed and shown to be effective in reducing $I_{th}$ and improving η$_{wp}$. Side stressors turned out to be a more efficient and easier way to increase η$_{wp}$ than using top and side stressors together. With the side stressors only and geometry optimizations, a η$_{wp}$ of 30.5%, and an $I_{th}$ of 50 mA ($J_{th}$ of 37 kA/cm$^2$) can be achieved with 1ns defect limited carrier lifetime of. These are tremendous improvements from the case without any stressors and geometry optimization, which has a η$_{wp}$ of 2.07% and an $I_{th}$ of 810 mA ($J_{th}$ of 300 kA/cm$^2$) respectively. With a longer defect-limited minority carrier lifetime (better material quality), the performance of Ge lasers can be further improved. With $\tau_{p,n} = 10\ ns$, an $I_{th}$ of 7.8 mA ($J_{th}$ of 5.8 kA/cm$^2$) and a wall-plug efficiency of 38.7% could be achieved at 121 mA and an output power of 33 mW. These results give a strong support to the Ge-on-Si laser technology and provide effective ways to improve the Ge laser performance.

## 8. Acknowledgments

This work was funded by the University of British Columbia. The authors thank Crosslight Software Inc. for providing the license of LASTIP$^{TM}$, Mr. Michel Lestrade from Crosslight Software for his assistance in LASTIP$^{TM}$ usage, and Mr. Xiyue Li from School of Electronic and Information Engineering, the South China University of Technology for helpful discussions.

## 9. Reference

[1] Siming Chen, Wei Li, Jiang Wu , Qi Jiang , Mingchu Tang , Samuel Shutts, Stella N. Elliott, Angela Sobiesierski, Alwyn J. Seeds , Ian Ross, Peter M. Smowton and Huiyun Liu * "Electrically pumped continuous-wave III-V quantum dot lasers on silicon" DOI: 10.1038/NPHOTON.2016.21

[2] Zhiping Zhou , Bing Yin and Jurgen Michel "On- chip light sources for silicon photonics." Light: Science & Applications (2015) 4, e358 doi:10.1038/lsa.2015.131

[3] Jifeng Liu, Xiaochen Sun, Dong Pan, Xiaoxin Wang, Lionel C. Kimerling, Thomas L. Koch, and Jurgen Michel "Tensile-strained, n-type Ge as a gain medium for monolithic laser integration on Si" Vol. 15, No. 18 / OPTICS EXPRESS 11272

[4] Jifeng Liu, Xiaochen Sun, Rodolfo Camacho-Aguilera, Lionel C. Kimerling, and Jurgen Michel "Ge-on-Si laser operating at room temperature" Vol. 35, No. 5 / OPTICS LETTERS

[5] Rodolfo E. Camacho-Aguilera, Yan Cai, Neil Patel, Jonathan T. Bessette, Marco Romagnoli, Lionel C. Kimerling, and Jurgen Michel1, "An electrically pumped germanium laser", Optics Express, Vol.20, No.10 (2012)

[6] Roman Koerner, Michael Oehme, Martin Gollhofer, Marc Schmid, Konrad Kostecki, Stefan Bechler, Daniel Widmann, Erich Kasper, and Joerg Schulze "Electrically pumped lasing from Ge fabry-perot resonator on Si." Optics Express Vol. 23, Issue 11, (2015)

[7] R. Geiger, S. Wirths , D. Buca and H. Sigg N. von den Driesch , Z. Ikonic , J.M. Hartmann , J. Faist , S. Mantl , D. Grützmacher "A Direct Band Gap GeSn Laser on Si" 10.1364/CLEO_SI.2015.SM3G.5

[8] Martyna Grydlik, Florian Hackl, Heiko Groiss, Martin Glaser, Alma Halilovic, Thomas Fromherz, Wolfgang Jantsch, Friedrich Schäffler, and Moritz Brehm "Lasing from Glassy Ge Quantum Dots in Crystalline Si" ACS Photonics, 2016, 3 (2), pp 298–303

[9] Birendra (Raj) Dutt, Devanand S. Sukhdeo, Donguk Nam , Boris M. Vulovic, 1 Ze Yuan, and Krishna C. Saraswat "Roadmap to an efficient germanium-on-silicon laser: Strain versus n-type doping" 10.1109/JPHOT.2012.2221692

[10] J. Liu, X. Sun, D. Pan, X. Wang, L. C. Kimerling, T. L. Koch, and J. Michel, "Tensile-strained, n-type Ge as a gain medium for monolithic laser integration on Si," Opt. Express, vol. 15, no. 18, pp.11272–11277, Aug.2007.

[11] J. Michel, R. E. Camacho-Aguilera, Y. Cai, N. Patel, J. T. Bessette, M. Romagnoli, R. Dutt, and L. Kimerling, : An electrically pumped Ge-on-Si laser, in Optical Fiber Communication Conference, OSA Technical Digest (Optical Society of America, 2012), paper PDP5A.6.

[12] Sukhdeo D S, Nam D, Kang J-H, Brongersma M L and Saraswat K C "Direct bandgap germanium-on-silicon inferred from 5.7% ⟨100⟩ uniaxial tensile strain" Photon. Res. / Vol. 2, No. 3 / June 2014

[13] G.Capellini et al., "Tensile Ge microstructures for lasing fabricated by means of a silicon complementary metal-oxide-semiconductor process" Opt Express. 2014 Jan 13;22(1):399-410

[14] Dirk Peschka, Marita Thomas, Annegret Glitzky, Reiner Nu¨rnberg, Michele Virgilio, Subhajit Guha, Thomas Schroeder, Giovanni Capellini, Thomas Koprucki "Robustness analysis of a device concept for edge-emitting lasers based on strained germanium" Opt Quant Electron (2016) 48:156

[15] Donguk Nam, Devanand Sukhdeo, Arunanshu Roy, Krishna Balram, Szu-Lin Cheng, Kevin Chih-Yao Huang, Ze Yuan, Mark Brongersma, Yoshio Nishi, David Miller, and Krishna Saraswat " Strained germanium thin film membrane on silicon substrate for optoelectronics" Vol. 19, Issue 27, pp. 25866-25872 (2011)

[16] Xiyue Li, Zhiqiang Li, Simon Li, Lukas Chrostowski and Guangrui (Maggie) Xia "Design Considerations of Biaxially Tensile-Strained Germanium-on-Silicon Lasers" 2016 Semicond. Sci. Technol. 31 065015

[17] S.Pidin, T.Mori, K.Inoue, S.Fukuta, N.Itoh, E.Mutoh, K.Ohkoshi, R.Nakamura, K.Kobayashi, K.Kawamura, T.Saiki, S.Fukuyama, S.Satoh, M.Kase, and K.Hashimoto "A Novel Strain Enhanced CMOS Architecture Using Selectively Deposited High Tensile And High Compressive Silicon Nitride Films" 10.1109/IEDM.2004.1419112

[18] Rodolfo E. Camacho-Aguilera, "Ge-on-Si LASER for Silicon Photonics", Ph. D dissertation, Department of Materials Science and Engineering, Massachusetts Institute of Technology, 2013.

[19] D.J Paul, "Si/SiGe heterostructures: from materials and physics to devices and circuits,"Semicond. Sci. Technol,vol.19, no.10, pp.R75-R108, Oct.2004.

[20] Karl Brunner, "Si/Ge nanostructures,"Rep. Prog. Phys., vol.65, no.1, pp. 27-72, Jan.2002.

[21] R.Camacho-Aguilera,Z.Han, Y.Cai,L.C.KimerlingandJ.Michel, "Direct band gap narrowing in highly doped Ge," Appl.Phys.Lett.,vol.102,no.15,p.152106,Apr.2013.

[22] Y. Cai, Z. Han, X. Wang, R.E. Camacho-Aguilera, L.C. Kimerling, J. Michel, and J. Liu, "Analysis of Threshold Current Behavior for Bulk and Quantum Well Germanium Laser Structures," IEEE Journal of Selected Topics in Quantum Electronic,vol. 19, no.4,p.1901009 , Jul.2013.

[23] J. Liu, X. Sun, D. Pan, X. Wang, L. C. Kimerling, T. L. Koch, and J. Michel, "Tensile-strained, n-type Ge as a gain medium for monolithic laser integration on Si," Opt. Express, vol. 15, no. 18, pp.11272–11277, Aug.2007.

[24] David S. Sukhdeo a , Shashank Gupta a , Krishna C. Saraswat a , Birendra (Raj) Dutt b,c , Donguk Nam d, "Impact of minority carrier lifetime on the performance of strained germanium light sources" Optics Communications 364 (2016) 233–237

[25] R. E. Camacho-Aguilera, "Monolithically-integrated Ge CMOS laser", SPIC Proceedings, vol.9010, Feb, 2014.

[26] R.Newman and W.W.Tyler, "Effect of impurities on free-hole infrared absorption in p-type germanium", Physical Review, vol.105,no.3, pp. 885-886, Feb.1957.

[27] Shun Lien Chuang "Physics of photonic devices" ISBN: 978-0-470-29319-5

[28] Geiger R, Frigerio J, Süess M J, Chrastina D, Isella G, Spolenak R, Faist J and Sigg H 2014 Excess carrier lifetimes in Ge layers on Si Appl. Phys. Lett. 104 062106

[29] Nam D, Kang J-H, Brongersma M L and Saraswat K C 2014 Observation of improved minority carrier lifetimes in highquality Ge-on-insulator using time-resolved photoluminescence Opt. Lett. 39 6205